\documentclass[a4paper,fleqn]{cas-dc}

\usepackage{algorithm}
\usepackage{algpseudocode}
\usepackage{tabularx, makecell, booktabs, multirow}

\usepackage[numbers,sort&compress]{natbib}
\usepackage{enumitem}
\usepackage{nomencl}

\makenomenclature

\usepackage{etoolbox}

\renewcommand{\nomgroup}[1]{%
  \item[\bfseries
  \ifstrequal{#1}{A}{Indices}{%
  \ifstrequal{#1}{B}{Sets}{%
  \ifstrequal{#1}{C}{Parameters}{%
  \ifstrequal{#1}{D}{Continuous Variables}{%
  \ifstrequal{#1}{E}{Binary Variables}{}}}}}]%
  \vspace{0.3\baselineskip}
}

\def\tsc#1{\csdef{#1}{\textsc{\lowercase{#1}}\xspace}}
\tsc{WGM}
\tsc{QE}

\begin{document}
\let\WriteBookmarks\relax
\def\floatpagepagefraction{1}
\def\textpagefraction{.001}

\shorttitle{Co-Optimization of GETs using Iterative McCormick Relaxation}
\shortauthors{Junseon Park et~al.}

\title [mode = title]{Co-Optimization of Network Topology and Variable Impedance Devices under Dynamic Line Ratings Using Iterative McCormick Relaxation}                      

\author[1]{Junseon Park}[]

\affiliation[1]{organization={Department of Safety Engineering, Pukyong National University},
                city={Busan},
                postcode={48513},
                country={Republic of Korea}}

\author[1]{Hyeongon Park}[]

\author[2]{Rahul K. Gupta}[]


\affiliation[2]{organization={Institute of Electrical and Micro Engineering, School of Engineering, EPFL},
                city={Lausanne},
                postcode={1015},
                country={Switzerland}}

\begin{abstract}
Grid Enhancing Technologies (GETs), such as Network Topology Optimization (NTO), Dynamic Line Rating (DLR), and Variable Impedance Devices (VIDs) are increasingly deployed in power transmission grids to address congestion and improve grid utilization. However, their growing adoption necessitates coordinated operation to fully exploit their benefits. This paper presents a unified co-optimization framework that integrates NTO, DLR, and VID to alleviate congestion and minimize operational costs. NTO is modeled using a node-breaker formulation to provide enhanced switching granularity, while DLR incorporates weather-dependent parameters, including wind speed and ambient temperature, to dynamically adjust transmission limits. The resulting formulation is a mixed-integer nonlinear program due to power flow constraints with bilinear VID terms, and discrete topology decisions. To improve tractability, we propose using a McCormick relaxation approach to linearize bilinear terms. Furthermore, we propose an iterative refinement of the McCormick relaxation scheme to enhance solution accuracy. The proposed framework is validated through detailed analyses on the IEEE 118-bus system, along with additional case studies on other benchmark systems to assess scalability. Results demonstrate that the proposed approach improves solution accuracy compared to standard McCormick relaxation, achieves solution quality close to the nonlinear formulation, and maintains computational efficiency.

\end{abstract}


\begin{highlights}
\item A unified co-optimization framework is developed for the coordinated operation of NTO, DLR, and VIDs.
\item An iterative McCormick relaxation method is proposed to efficiently handle bilinear terms arising from VID operation.
\item Benchmark results show that the proposed method maintains high solution accuracy and computational efficiency across different test systems.
\end{highlights}

\maketitle

\nomenclature[A1]{$t$}{Time index}
\nomenclature[A2]{$b$}{Substation index}
\nomenclature[A3]{$l$}{Line index}
\nomenclature[A4]{$e$}{Line-end index ($e \in \{fr, to\}$)}
\nomenclature[A5]{$i$}{Busbar index ($i \in \{1, 2\}$)}
\nomenclature[A6]{$g$}{Generator index}
\nomenclature[A7]{$d$}{Load index}

\nomenclature[B1]{$\mathcal{T}$}{Set of time periods}
\nomenclature[B2]{$\mathcal{B}$}{Set of Substations}
\nomenclature[B3]{$\mathcal{L}$}{Set of transmission lines}
\nomenclature[B4]{$\mathcal{G}$}{Set of generators}
\nomenclature[B5]{$\mathcal{D}$}{Set of loads}
\nomenclature[B6]{$\mathcal{G}_b$}{Set of generators at substation $b$}
\nomenclature[B7]{$\mathcal{D}_b$}{Set of loads at substation $b$}
\nomenclature[B8]{$\mathcal{L}_b^{\mathrm{fr}}$}{Set of lines connected whose directions of power flow are from substation $b$}
\nomenclature[B9]{$\mathcal{L}_b^{\mathrm{to}}$}{Set of lines connected whose directions of power flow are to substation $b$}

\nomenclature[C]{$\bar{b}_{l}$}{Nominal susceptance of line $l$ [S]}
\nomenclature[C]{$T_{\mathrm{avg}}$}{Average temperature of conductor [$^\circ$C]}
\nomenclature[C]{$R(T_{\mathrm{avg}})$}{Conductor resistance at $T_{\mathrm{avg}}$ [$\Omega$]}
\nomenclature[C]{$mC_p$}{Total heat capacity of conductor [J/(m$\cdot^\circ$C)]}
\nomenclature[C]{$q_c$}{Convection heat loss rate per unit length [W/m]}
\nomenclature[C]{$q_s$}{Heat gain rate from sun [W/m]}
\nomenclature[C]{$q_r$}{Radiated heat loss rate per unit length [W/m]}
\nomenclature[C]{$\alpha_l^{t}$}{Ratio of allowable current under dynamic and static line ratings}
\nomenclature[C]{$\theta^{\max}$}{Maximum allowed voltage angle [rad]}
\nomenclature[C]{$P_g^{\min}$}{Minimum generation of generator $g$ [MW]}
\nomenclature[C]{$P_g^{\max}$}{Maximum generation of generator $g$ [MW]}
\nomenclature[C]{$\hat{P}_{b,d}^t$}{Forecasted load $d$ at substation $b$ at time $t$ [MW]}
\nomenclature[C]{$I$}{Conductor line rating [A]}
\nomenclature[C]{$M$}{A sufficiently large number}
\nomenclature[C]{$r$}{Susceptance change rate}
\nomenclature[C]{$c_{g,k}$}{Cost coefficient of generator $g$, $k \in \{0,1\}$}
\nomenclature[C]{$\mathrm{VOLL}$}{Value of Lost Load [$\$/\mathrm{MWh}$]}
\nomenclature[C]{$\overline{N}_b$}{Maximum total number of bus splitting actions allowed over the scheduling horizon}
\nomenclature[C]{$\overline{N}_l$}{Maximum total number of line switching actions allowed over the scheduling horizon}
\nomenclature[C]{$RR_g^{up}$}{Ramp-up limit of generator $g$ [MW]}
\nomenclature[C]{$RR_g^{dn}$}{Ramp-down limit of generator $g$ [MW]}

\nomenclature[D]{$P_l^t$}{Power flow on line $l$ at time $t$ [MW]}
\nomenclature[D]{$P_{g,i}^t$}{Generation power connected to busbar $i$ at time $t$ [MW]}
\nomenclature[D]{$P_{b, d,i}^t$}{Load $d$ connected to busbar $i$ in substation $b$ at time $t$ [MW]}
\nomenclature[D]{$P_{l,e,i}^t$}{Power flow on line $l$ at end $e$ connected to busbar $i$ at time $t$ [MW]}
\nomenclature[D]{$\theta_{l,e}^t$}{Voltage angle of line $l$ at end $e$ at time $t$ [rad]}
\nomenclature[D]{$\theta_{b,i}^t$}{Voltage angle at busbar $i$ in substation $b$ at time $t$ [rad]}
\nomenclature[D]{$\theta_{l,e,i}^t$}{Voltage angle at busbar $i$ associated with end $e$ of line $l$ at time $t$ [rad]}
\nomenclature[D]{$\Delta b_l^t$}{Susceptance change of line $l$ at time $t$ [S]}
\nomenclature[D]{$C^{\mathrm{Gen}}$}{Total generation cost [$\$$]}
\nomenclature[D]{$C^{\mathrm{LS}}$}{Load shedding cost based on VOLL [$\$$]}
\nomenclature[D]{$C^{\mathrm{Obj}}$}{Total operational cost [$\$$]}
\nomenclature[D]{$b_l^t$}{Susceptance of line $l$ at time $t$}
\nomenclature[D]{$\Delta \theta_l^t$}{Difference of voltage angle of line $l$ at end $e$ at time $t$ [rad]}
\nomenclature[D]{$w_{l}^t$}{Auxiliary variable representing $\Delta b_{l}^t \cdot \Delta \theta_{l}^t$}
\nomenclature[E]{$h_b^t$}{Binary variable for connection status of busbars in substation $b$ at time $t$(0: disconnected, 1: connected)}
\nomenclature[E]{$h_g^t$}{Binary variable for busbar connection of generator at time $t$(0: busbar 1, 1: busbar 2)}
\nomenclature[E]{$h_d^t$}{Binary variable for busbar connection of load at time $t$(0: busbar 1, 1: busbar 2)}
\nomenclature[E]{$h_{l,e}^t$}{Binary variable for busbar connection of line $l$ end $e$ at time $t$ (0: busbar 1, 1: busbar 2)}
\nomenclature[E]{$h_l^t$}{Binary variable for switching status of line $l$ at time $t$(0: open, 1: closed)}

\printnomenclature

\section{Introduction}
The bulk power system is undergoing a massive transformation driven by the large-scale integration of intermittent renewable energy resources (RERs), inverter-based resources (IBRs) \cite{lin2022pathways}, and the rapid growth of data center loads \cite{DOE_report_RA}. These developments introduce new operational challenges such as power flow congestion, voltage quality and stability issues \cite{liang2016emerging, NERC_report}, making it increasingly difficult for the power system utilities to ensure power system reliability. Conventional solutions to these challenges are to expand transmission capacity through new line construction or infrastructure upgrades \cite{lee2006transmission, de2008transmission}, deploy energy storage systems (ESS) \cite{denholm2010role, piansky2025optimizing}, etc. However, these approaches are often capital-intensive and time-consuming, requiring several years for planning, permitting, and commissioning.

In recent years, several Grid Enhancing Technologies (GETs) have emerged as promising alternatives to improve network flexibility and mitigate congestion issues \cite{DOE_report, mirzapour2024grid, su2025grid}. These include Network Topology Optimization (NTO) \cite{bacher2007network, 7038226, xiao2018power}, Dynamic Line Rating (DLR) \cite{xiao2018power, DOE_report_DLR}, and Variable Impedance Devices (VIDs), also referred to as Smart Wire Devices (SWDs) \cite{kreikebaum2010smart, sahraei2016computationally}.

\begin{figure}
	\centering
    \includegraphics[width=\columnwidth]{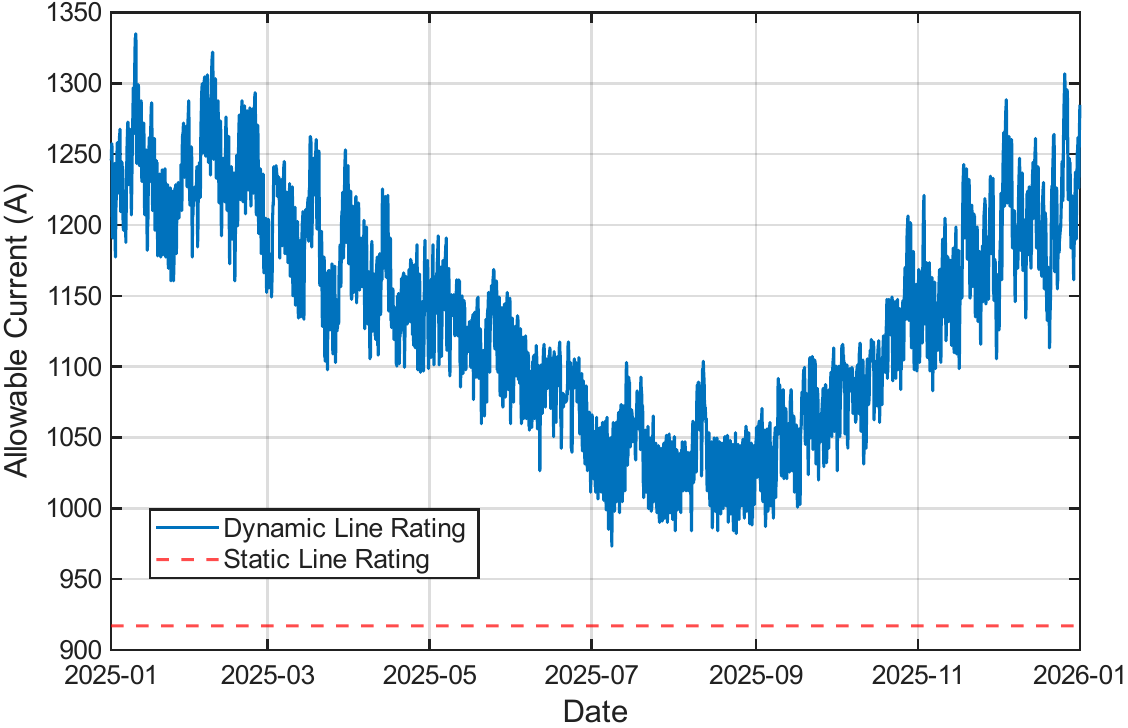}
    \vspace{-1.5em}
	\caption{Allowable current under DLR and static line rating (SLR) in South Korea. The SLR is determined based on conservative worst-case ambient conditions defined by standard practice, while the DLR is computed using meteorological data\cite{KMA_portal}.}
	\label{fig:dlr_real_Data}
\end{figure}

NTO seeks to optimize both the network topology and the breaker configurations, often represented by the node-breaker model, to enhance operational flexibility. Prior studies \cite{xiao2018power} have shown that NTO can significantly reduce operational costs compared to fixed-topology systems. Moreover, DLR exploits the dynamically changing thermal limits of transmission lines based on real-time ambient weather conditions, allowing operators to utilize increased line capacity during favorable cooling conditions such as low temperatures and high wind speeds \cite{DOE_report_DLR}. However, as shown in Figure~\ref{fig:dlr_real_Data}, the allowable current exhibits significant temporal variability, which may lead to operational challenges when not properly coordinated with other grid control mechanisms.

Meanwhile, VIDs (or SWDs) can modulate the line impedance, effectively rerouting power flows from overloaded lines to underutilized ones \cite{sahraei2016computationally}. The operation of VIDs has been investigated for several objectives, including congestion management \cite{nikoobakht2018smart}, cost minimization, and wind power curtailment reduction \cite{nasri2014minimizing}. In addition, \cite{ziaee2016stochastic, sahraei2015fast, sang2017stochastic} developed stochastic planning and operational frameworks for the deployment of VIDs under uncertainty. Recognizing the potential of these technologies, the U.S. Department of Energy (DOE) has recently emphasized the importance of GET deployment to enhance grid reliability and flexibility \cite{DOE_report}. However, most existing methods address these technologies independently, without considering their interactions or coordination in congestion management. To address this gap, this work proposes a co-optimization framework that jointly models and coordinates NTO, DLR, and distributed VIDs, enabling synergistic operation to alleviate transmission congestion and improve overall system efficiency.

In the existing literature, the optimization of GETs across various operational time horizons has been widely investigated. Recent studies have examined the combined application of Optimal Transmission Switching (OTS) and DLR technologies \cite{xiao2016bulk, wu2025dynamic}. These works have demonstrated that integrating OTS and DLR schemes can significantly enhance renewable energy hosting capacity and improve overall system reliability. In \cite{7038226}, a DLR along with NTO framework, combining OTS and bus-bar splitting, was proposed, showing substantial improvements in system reliability and notable reductions in operational costs.
Moreover, incorporating VIDs into operational planning introduces additional nonlinearities in both the optimal power flow (OPF) and switching problems, as VIDs are typically modeled through variable line impedance \cite{sahraei2016computationally, nikoobakht2018smart}. 

Overall, the existing body of work has predominantly focused on the individual optimization of specific GETs, with limited attention to their coordinated or co-optimized operation. For instance, while \cite{7038226} explored the integration of DLR within an NTO framework, and \cite{ziaee2016stochastic} considered VID deployment alongside generator dispatch scheduling. 
Consequently, there is a clear research gap in developing unified frameworks that enable the coordinated operation of multiple GETs to fully leverage their combined flexibility and enhance system performance. 

In this work, we propose a co-optimization framework for the coordinated operation of multiple GETs. Our framework simultaneously considers three complementary technologies: NTO, DLR, and VIDs, to exploit their synergistic capabilities in improving power system flexibility and operational efficiency.
\textit{First}, the network topology and bus-bar splitting are modeled using the node-breaker representation, which enables detailed switching decisions while maintaining network connectivity and operational feasibility. \textit{Second}, VIDs are incorporated to provide additional flexibility by allowing controlled adjustments of line reactances within predefined operating limits, thereby redistributing power flows and alleviating congestion. \textit{Third}, DLR is modeled as functions of real-time or forecasted weather conditions, primarily wind speed and ambient temperature, enabling adaptive utilization of transmission capacity based on prevailing environmental factors. 

The proposed formulation to allow flexible adjustment of line reactance results in a bilinear optimization problem, where constraints are expressed as the product of variables. This makes it difficult to find the optimal solution. To overcome the challenge, we employ a McCormick Relaxation method to relax the bilinear constraints. 
While the McCormick Relaxation method reduces computation time, it has the drawback of lower accuracy. To tackle this trade-off, we develop  an Iterative McCormick Relaxation, that solves the problem by repeating McCormick Relaxation with smaller intervals. Applying this method reduces the computational burden associated with traditional bilinear constraints and improves accuracy compared to McCormick Relaxation. 

The key contribution of this work lies in the development of a co-optimization scheme that jointly considers NTO, DLR, and VID technologies within a single, integrated decision-making framework and its linearization using the proposed Iterative McCormick Relaxation scheme. This unified approach enables a systematic assessment of the interactions and combined benefits of multiple GETs under diverse operating and weather conditions. Through this framework, we can evaluate how coordinated control of these technologies enhances network flexibility, increases renewable energy accommodation, and improves overall system reliability, outcomes that are difficult to achieve when each technology is optimized independently.

The paper is organized as follows. Section~\ref{sec:methodology} presents the transmission network model and the GET models used in this study, including NTO, DLR, and VIDs. Section~\ref{sec:co-optimization} presents the co-optimization framework for the coordination of the multiple GETs. Section~\ref{sec:solution_approach} presents the solution methodologies for handling the bilinear constraints associated with VID, including the original nonlinear formulation, standard McCormick relaxation, and the proposed iterative McCormick relaxation. Section~\ref{sec:Simulation} presents the simulation setup and results on IEEE test cases, and Section~\ref{sec:conclusion} concludes the main contribution of the presented work.

\section{Modeling Grid Enhancing Technologies}
\label{sec:methodology}
\begin{figure}
	\centering
    \includegraphics[width=\columnwidth]{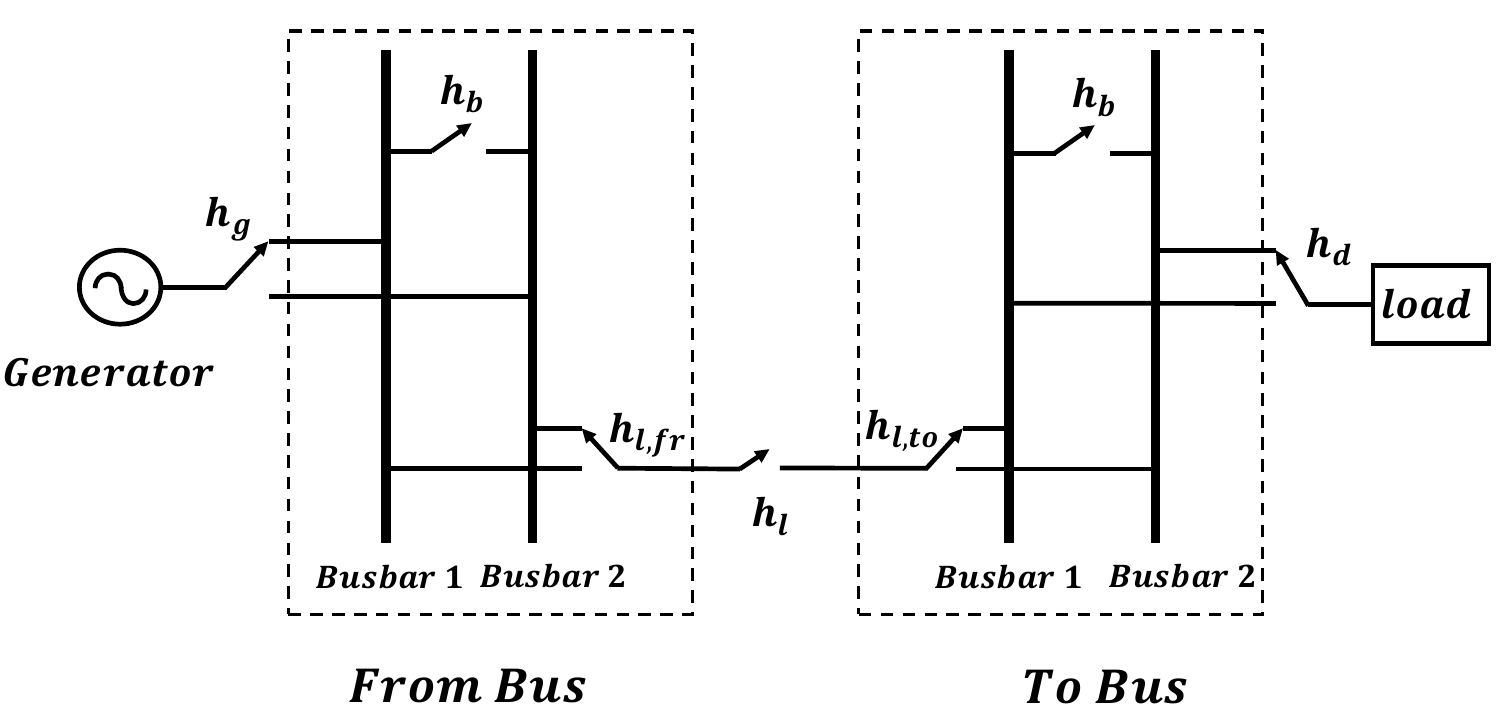}
    \vspace{-1.5em}
	\caption{Generalized substation model.}
	\label{FIG:1}
\end{figure}
\subsection{Network Topology Optimization Model}
NTO involves optimizing the configuration of the power network through optimal transmission switching and by leveraging existing switching elements within substations, such as circuit breakers (CBs). The substation CBs enable reconfiguration of substation connections to the rest of the power system, thereby influencing the operational flexibility, reliability, and security of the grid. 

In this work, the power network is modeled using a node-breaker representation at each bus~\cite{7038226, xiao2018power}. Figure~\ref{FIG:1} shows the generalized substation model with line switching. As shown in the figure, the node-breaker provides additional operational flexibility by allowing the busbars at the ``from'' and ``to'' buses can be either connected or separated. In addition, generators and loads can be assigned to either busbars, thereby enabling more flexibility network reconfiguration. The NTO model using busbar binary variables $h_b^t$ is given as follows
\begin{subequations}    
\label{eq:NTO_equations}
\begin{align}
& -\theta^{\max} (1 - h_b^t) \leq \theta_{b, 1}^t - \theta_{b, 2}^t \leq \theta^{\max} (1 - h_b^t) \label{max_voltage}
\end{align}
for all $b \in \mathcal{B}$ and $t \in \mathcal{T}$. In \eqref{max_voltage}, when busbars are merged ($h_b^t=1$), the difference in voltage angle phase between busbars must be identical, and when split, it must not exceed the maximum angle phase. 

The generators connected to each busbar can be modeled using binary variables $h_g^t$, and can be expressed as
\begin{align}
(1-h_g^t)P_g^{\min} \leq P_{g, 1}^t &\leq (1 - h_g^t)P_g^{\max} &&\forall g, t \label{Pgmax_1}\\
h_g^tP_g^{\min} \leq P_{g, 2}^t &\leq h_g^tP_g^{\max} &&\forall g, t \label{Pgmax_2}\\ 
\intertext{where, \eqref{Pgmax_1} and \eqref{Pgmax_2} models when a generator is connected to busbar 1 or at busbar 2, and vice versa. In addition, the ramping limits of generators are modeled as}
P_g^t - P_g^{t-1} &\leq RR_g^{up} &&\forall g, t \\
P_g^{t-1} - P_g^{t} &\leq RR_g^{dn} &&\forall g, t \\
\intertext{
Similarly, the connection of load at each busbar is modeled using the binary variable $h_d^t$ as}
0 \leq P_{b,d, 1}^t &\leq (1 - h_d^t)\hat{P}_{b,d}^t &&\forall b, t \label{Pdmax_1}\\
0 \leq P_{b,d, 2}^t &\leq h_d^t \hat{P}_{b,d}^t &&\forall b, t \label{Pdmax_2}
\end{align}
The power flow limits and line switching constraints, applied to all $l \in \mathcal{L}$, $e \in \{fr,to\}$, and $t \in \mathcal{T}$, are expressed as
\begin{align}
  -(1 - h_{l, e}^t)P_l^{\max, t} &\leq P_{l, e, 1}^t \leq (1 - h_{l, e}^t)P_l^{\max, t}  \label{Plmax_1}\\
 - h_{l, e}^tP_l^{\max, t} &\leq P_{l, e, 2}^t \leq h_{l, e}^tP_l^{\max, t}  \label{Plmax_2}\\
 -h_{l}^tP_l^{\max, t} &\leq P_{l, e, 1}^t \leq h_{l}^tP_l^{\max, t} \label{hl_max}\\
 h_{l, e}^t &\leq h_{l}^t \label{hl}\\
 P_l^t &= P_{l, e, 1}^t + P_{l, e, 2}^t  \label{sumPl}
\end{align} 
Here, $P_l^{\max,t}$ denotes the transmission line capacity at time $t$. Under static line rating (SLR), this capacity remains constant over time and is denoted by $P_l^{\max}$, whereas under DLR, it becomes time-varying as $P_l^{\max,t} = \alpha_l^t \cdot P_l^{\max}$.

Constraints \eqref{Plmax_1} and \eqref{Plmax_2} define the transmission line constraints: when a line is connected to busbar 1, the power flow through the other busbar is forced to zero, and vice versa. Constraints \eqref{hl_max} and \eqref{hl} specify the line status and enforce zero power flow when the line is open. Constraint \eqref{sumPl} enforces power conservation by requiring that the total line flow equals the sum of the flows associated with the two busbars at each line end. 

We model the transmission network using the DC power flow approximation \cite{stott2009dc}, which assumes that voltage angle differences are small, line resistances are negligible compared with reactances, and voltage magnitudes are close to 1~per unit. The line switching constraints are expressed using the big-$M$ method as
\begin{align}
   & -M(1 - h_l^t) 
   \leq  b_{l}^t(\theta_{l, fr}^t - \theta_{l, to}^t) - P_l^t \leq M(1 - h_l^t) \label{Pf_NTO}\\
   & -h_{l, e}^t\theta^{\max} 
   \leq  \theta_{l, e}^t - \theta_{l, e, 1}^t \leq h_{l, e}^t\theta^{\max} \label{theta_max_1}\\
   & -(1 - h_{l, e}^t)\theta^{\max} 
   \leq \theta_{l, e}^t - \theta_{l, e, 2}^t \leq (1 - h_{l, e}^t)\theta^{\max} \label{theta_max_2}
\end{align}
for all $l \in \mathcal{L}$, $e \in \{fr,to\}$, and $t \in \mathcal{T}$.
Here, $b_l^t = \bar{b}_l$, i.e., the line susceptance is fixed when impedance control is not considered.

The binary variables for busbars, generators, loads, and line ends are linked through \eqref{hg1}--\eqref{hle1}. When busbars at a substation are connected (i.e., $h_b^t=1$), the binary variables associated with generator, load, and line-end connections become redundant as they are connected to busbar 1 by default. This is expressed by following constraints.
\begin{align}
& \qquad h_b^t + h_g^t \leq 1 &&\forall b, t, g \label{hg1}\\
& \qquad h_b^t + h_d^t \leq 1 &&\forall b, t, d \label{hd1}\\
& \qquad h_b^t + h_{l, e}^t \leq 1 &&\forall b, t, e, l \label{hle1}
\end{align}

We also impose a constraint on the total number of bus splitting and line switching operations over the entire time horizon as follows.
\begin{align}
\qquad \sum_{t \in \mathcal{T}} \sum_{b \in \mathcal{B}} (1 - h_b^t) \leq \overline{N} _b \label{spliting_time} \\
\qquad \sum_{t \in \mathcal{T}} \sum_{l \in \mathcal{L}} (1 - h_l^t) \leq \overline{N}_l \label{connecting_time}
\end{align}

Constraint \eqref{spliting_time} limits the total number of bus splitting actions over the time horizon, while \eqref{connecting_time} limits the total number of line switching actions.

Finally, the power balance constraints at each busbar are expressed by \eqref{balance1} and \eqref{balance2}, for all $b \in \mathcal{B}$ and $t \in \mathcal{T}$, ensuring that the net power injection and net line flow are balanced.

\begin{align}
 \sum_{g \in G_b}P_{g, 1}^t - \sum_{d \in D_b}P_{b, d, 1}^t - \sum_{l \in LF_b}P_{l}^t + \sum_{l \in LT_b}P_{l}^t = 0 \label{balance1}\\
 \sum_{g \in G_b}P_{g, 2}^t - \sum_{d \in D_b}P_{b, d, 2}^t - \sum_{l \in LF_b}P_{l}^t + \sum_{l \in LT_b}P_{l}^t = 0 \label{balance2}
\end{align}
\end{subequations}

\subsection{Dynamic Line Rating Model}
We also consider the DLR capability in our optimization scheme as one of the GETs. Compared to SLR, which operates by fixing transmission capacity, DLR enables more economical power system operation by providing higher transmission capacity. 

The DLR is modeled based on weather data and follows the IEEE Standard 738-2023 \cite{10382442}. With notations defined in the nomenclature, the dynamic thermal balance of a conductor can be expressed as
\begin{subequations}  
\begin{align}
& \frac{dT_{avg}}{dt} = \frac{1}{mC_{p}}[R(T_{avg})I^{2} + q_{s} - q_{c} - q_{r}] \label{DLR_Non}
\end{align}

As the ambient temperature rises, $q_{c}$ and $q_{r}$ decrease, reducing cooling efficiency and current capacity. In contrast, higher wind speed enhances $q_{c}$, which cools the conductor more effectively and increases the allowable current. Overall, ambient temperature and wind speed are the dominant weather factors determining the dynamic line rating. For the hourly scheduling problem considered in this study, the conductor temperature is assumed to reach a steady state within each time interval. Therefore, following \cite{nick2015security}, the dynamic term $\frac{dT_{avg}}{dt}$ is neglected, and the steady-state condition of the thermal balance yields,
\begin{align}
I = \sqrt{\frac{q_{c} + q_{r} - q_{s}}{R(T_{avg})}} \label{DLR_lin}
\end{align}
\end{subequations}
providing simplified expression for allowable current as function of weather conditions \cite{lai2023optimisation}. To model the DLR effect in optimization, we approximate in \eqref{Plmax_1}--\eqref{sumPl} as
\begin{align}
\label{DLR_Cons}
P_l^{\max,t} = \alpha_l^{t} \, P_l^{\max}
\end{align}
where $\alpha_l^{t}$ is a scaling factor derived from the ratio between the allowable currents under DLR and SLR.
To determine $\alpha_l^{t}$, the allowable current under DLR, denoted by $I^{DLR}$, is first computed from \eqref{DLR_lin} using forecasted ambient temperature and wind speed. We also compute the reference current under SLR, $I^{SLR}$, which is obtained under fixed worst-case ambient conditions (40°C and 0.5 m/s wind speed). The scaling factor $\alpha_l^{t}$ is defined as the ratio between these two currents, and is used to adjust the baseline transmission capacity accordingly. A higher temperature or lower wind speed results in a smaller $\alpha_l^{t}$, whereas a lower temperature or higher wind speed yields a larger $\alpha_l^{t}$.

\subsection{Variable Impedance Devices Model}
VIDs are a type of flexible AC transmission system (FACTS). They are modeled 
are several kinds of FACTS devices, including Thyristor Controlled Series Compensators (TCSCs), Static Synchronous Series Compensators (SSSCs), and Unified Power Flow Controllers (UPFCs) and are typically modeled through reactances that can be varied by pre-defined limits. The variable impedance/reactance allows flexibility in the transmission network to re-route the power-flows and help avoid congestion and reduce costs. It can be mathematically modeled as \cite{sahraei2016computationally, nikoobakht2018smart, 8403390}
\begin{subequations}
\label{eq:VID}
\begin{align}
    b_l^t & = \bar{b}_{l} + \Delta b_l^t \\
    -r \bar{b}_l & \leq \Delta b_l^t \leq r \bar{b}_l \quad \forall l, t \label{rangeb}
\end{align}
\end{subequations}
where, $\bar{b}_l$ denotes the nominal susceptance and $\Delta b_l^t$ is a variable allowing deviations in the line susceptance is bounded by a factor $r \leq 1$ of nominal susceptance $\bar{b}_l$.

\section{Co-optimization Framework}
\label{sec:co-optimization}
We formulate a co-optimization problem that jointly coordinates the operation of multiple GETs, aiming to minimize the total operational cost while satisfying the respective constraints of each technology. 



The proposed optimization aims to minimize the total system operational cost, which consists of (i) generation cost $(C^{Gen})$ and (ii) load-shedding cost $(C^{LS})$.
Each cost component is defined as follows.
\begin{subequations}

The generation costs are modeled as a linear function of the total power output. Let $P_g^t$ be the total power output of generator $g$ at time $t$, defined as the sum of its outputs at the two split busbars, i.e., $P_g^t = P_{g,1}^t + P_{g,2}^t$. Accordingly, the generation cost is expressed as follows:
\begin{align}
C^{\mathrm{Gen}} =
\sum_{t \in \mathcal{T}} \sum_{g \in \mathcal{G}}
\left[
c_{g,1}P_g^t + c_{g,0}
\right]
\label{gen_cost}
\end{align}

This linear formulation can also be extended to represent piecewise linear approximations of quadratic generation cost curves.

The load shedding cost is determined based on the amount of unmet load. The load supplied at bus $b$ for load $d$ at time $t$ is given by $P_{b,d}^t = P_{b,d,1}^t + P_{b,d,2}^t$. Load shedding is penalized using the value of lost load (VOLL), and the total cost is expressed as
\begin{align}
C^{\mathrm{LS}} = \mathrm{VOLL} \sum_{t \in \mathcal{T}} \sum_{b \in \mathcal{B}}
\sum_{d \in \mathcal{D}_b}
\left( P_{b,d}^t - \hat{P}_{b,d}^t \right)
\label{load_curtail}
\end{align}
\end{subequations}

The final optimization for coordinating the operation of the NTO, DLR and VID can be formulated as 
\begin{equation}
\label{eq:optimization_problem_final}
\begin{aligned}
\min \quad &  C^{Obj} \;=\; C^{Gen} + C^{LS}, \\
\text{subject to} \quad 
&\eqref{eq:NTO_equations}, \eqref{DLR_Cons},\eqref{eq:VID}.
\end{aligned}
\end{equation}

\section{Solution Methodologies}
\label{sec:solution_approach}
The optimization problem in \eqref{eq:optimization_problem_final} is non-linear due to the bilinear constraints associated with VID when \eqref{eq:VID} is substituted into \eqref{Pf_NTO}. Since the methodology used to handle these non-linearities plays a critical role in determining the trade-off between numerical precision and computational efficiency, we investigate different solution strategies to identify the most effective approach.

To evaluate the effectiveness of the proposed approach, we consider three distinct solution methodologies: \textsf{Nonlinear}, \textsf{Static McCormick}, and \textsf{Iterative McCormick}. This framework enables a systematic assessment of numerical precision and computational tractability.

\subsection{\textsf{Nonlinear} Approach}
The \textsf{Nonlinear} approach solves the original bilinear formulation in \eqref{eq:optimization_problem_final} directly using an optimization solver capable of handling non-convexity. This method serves as a benchmark for evaluating solution quality, but it can be computationally intensive, particularly as the network size or the number of topology decision variables increases.

\subsection{\textsf{Static McCormick} Approach}
The standard McCormick relaxation is applied in a non-iterative manner and is referred to as \textsf{Static McCormick} in this paper. The \textsf{Static McCormick} approach relaxes the bilinear terms associated with VID by introducing auxiliary variables and McCormick envelopes. Specifically, we define
$w_l^t = \Delta b_l^t \Delta\theta_l^t$, where 
$\Delta\theta_l^t = \theta_{l,fr}^t - \theta_{l,to}^t$. Given the bounds 
$\Delta b_l^t \in [\Delta b_l^{\min}, \Delta b_l^{\max}]$ and 
$\Delta\theta_l^t \in [\Delta\theta_l^{\min}, \Delta\theta_l^{\max}]$, the following McCormick relaxation constraints are imposed for all $l$ and $t$:
\begin{subequations} 
\label{eq:mccormick}
\begin{align}
w_{l}^t &\geq \Delta b_l^{\min}\Delta\theta_{l}^t 
+ \Delta b_{l}^t\Delta\theta_l^{\min} 
- \Delta b_l^{\min}\Delta\theta_l^{\min} \label{mc_1}\\
w_{l}^t &\geq \Delta b_l^{\max}\Delta\theta_{l}^t 
+ \Delta b_{l}^t\Delta\theta_l^{\max} 
- \Delta b_l^{\max}\Delta\theta_l^{\max} \label{mc_2}\\
w_{l}^t &\leq \Delta b_l^{\min}\Delta\theta_{l}^t 
+ \Delta b_{l}^t\Delta\theta_l^{\max} 
- \Delta b_l^{\min}\Delta\theta_l^{\max} \label{mc_3}\\
w_{l}^t &\leq \Delta b_l^{\max}\Delta\theta_{l}^t 
+ \Delta b_{l}^t\Delta\theta_l^{\min} 
- \Delta b_l^{\max}\Delta\theta_l^{\min}. \label{mc_4}
\end{align}
\end{subequations}

By replacing the bilinear term with $w_l^t$ and imposing \eqref{eq:mccormick}, the original non-convex constraints are relaxed into a set of linear inequalities. As a result, the optimization problem is reformulated as a mixed-integer linear programming (MILP) problem, improving computational tractability.

\subsection{\textsf{Iterative McCormick} Approach}
The proposed \textsf{Iterative McCormick} approach is designed to reduce the relaxation gap while preserving the computational tractability of an MILP framework. As detailed in Algorithm~\ref{alg:alg1}, this approach refines the McCormick relaxation through incremental updates of the variable bounds. By progressively tightening the feasible region, \textsf{Iterative McCormick} improves solution accuracy without directly solving the original bilinear non-convex formulation.

\begin{algorithm}[!htbp]
\caption{Iterative McCormick Relaxation with Incremental $r$ Steps}
\begin{algorithmic}[1]
\Require Target relaxation level $r_{\text{final}}$, step size $\Delta r$, initial susceptances ${\bar{b}_l}$
\State Initialize $\hat{b}_l^t \gets \bar{b}_l$ for all lines $l$
\While{$r < r_{\text{final}}$}
    \State Solve \eqref{eq:optimization_problem_final} using standard McCormick relaxation \eqref{eq:mccormick} for  $r = \Delta r$
    \State Compute susceptance update $\Delta b_l^t$
    \State Update susceptance: $\hat{b}_l^t \gets b_l^t + \Delta b_l^t$ for all $l$
    \State $r \gets r + \Delta r$
\EndWhile
\State \textbf{return} final solution at $r_{\text{final}}$
\end{algorithmic}
\label{alg:alg1}
\end{algorithm}

\section{Numerical Experiments}
\label{sec:Simulation}
\subsection{Simulation Setup}

\subsubsection{Test Systems and Data}

We validate the proposed framework on several transmission test systems of varying scales. First, detailed results and in-depth analyses are presented for the IEEE 118-bus system. To further demonstrate the scalability and robustness of the proposed McCormick-based approach, we then validate the scheme on additional test cases, including \texttt{case24\_ieee\_rts}, \texttt{pglib\_case588}, and \texttt{case1354pegase}. All network data are obtained from the MATPOWER test \cite{zimmerman2010matpower}.

For the load data, we use a 24-hour profile with one-hour intervals from \cite{9163265}. For each test system, the load at each time period is determined by scaling the nominal bus load using the normalized hourly load profile. To model the DLR constraints, ambient temperature and wind speed data are obtained from the Austin, Texas area in 2023 \cite{NOAA_NCEI_CDO}. These weather variables are used to compute the time-varying thermal rating coefficient $\alpha_{l}^{t}$ in \eqref{DLR_Cons}.

\subsubsection{Parameter Settings}
In the simulations, the minimum output of all generators is set to zero in order to avoid situations where a generator's minimum output would exceed the load at its corresponding bus. For simulations with VIDs, the susceptance change factor $r$ in \eqref{rangeb} is set to 0.1, allowing a $\pm 10\%$ adjustment of the nominal line susceptance. The VOLL is set to 2{,}000~\$/MWh. To ensure a consistent baseline for comparison, the SLR and DLR cases use the same parameter settings except for the line rating values.

\subsubsection{Implementation and Solution Methods}
The optimization models were implemented in MATLAB and solved using Gurobi 12.0.3\footnote{\url{https://www.gurobi.com}} as the primary solver. The simulations were performed on a system equipped with an Apple M4 Pro chip (14-core CPU) and 24~GB of unified memory. Following the solution methodologies described in Section~\ref{sec:solution_approach}, we compare three approaches \textsf{Nonlinear}, \textsf{Static McCormick}, and \textsf{Iterative McCormick}. For the proposed \textsf{Iterative McCormick} approach, the incremental step size $\Delta r$ is set to $5\%$. 

\subsubsection{Simulation Scenarios}

To examine how NTO affects the synergistic benefits of VID and DLR, we consider two simulation scenarios: fixed topology and optimized topology.

\begin{enumerate}[label=(\roman*)]
    \item \textbf{Fixed topology:} 
    The network configuration is restricted to its nominal state, while the operations of DLR and VID are optimized. In this scenario, the topology-related binary variables ($h_l^t, h_b^t, h_g^t$) in \eqref{eq:optimization_problem_final} are enforced to $1$ to represent the base topology (i.e., no line switching or bus splitting). 

    \item \textbf{Optimized topology:} 
    The network topology is co-optimized with DLR and VID. Due to the computational complexity introduced by the topology-related binary decision variables, the optimization process is terminated upon reaching a 0.5\% optimality gap or a maximum execution time of 3{,}600 seconds. 
\end{enumerate}

In both scenarios, a congested grid environment is established by imposing a power-flow limit on each line equal to 80\% of its base-case loading. Furthermore, considering the high investment costs associated with GETs \cite{pohl2024coordination}, we evaluate partial deployment scenarios where DLR and VID are applied only to a prioritized subset of lines. The deployment priority is determined based on the \textit{average utilization rate} of each line, calculated from a base-case optimal power flow without VID, DLR, and NTO. Lines with higher average utilization rates are selected first for DLR and VID deployment. To assess the marginal impact of GETs, the deployment level is incrementally increased as a percentage of the total number of lines in the system.

\subsection{Simulation Results}

\subsubsection{Fixed Topology}
For the fixed-topology scenario, the original bilinear formulation can be solved directly using the \textsf{Nonlinear} approach. Therefore, we first present the results obtained from the \textsf{Nonlinear} approach as a reference for assessing the cost reduction achieved by DLR and VID. These results also serve as a benchmark for evaluating the accuracy and computational performance of the McCormick-based approaches.

\begin{figure}
    \centering
    \includegraphics[width=1.1\linewidth]{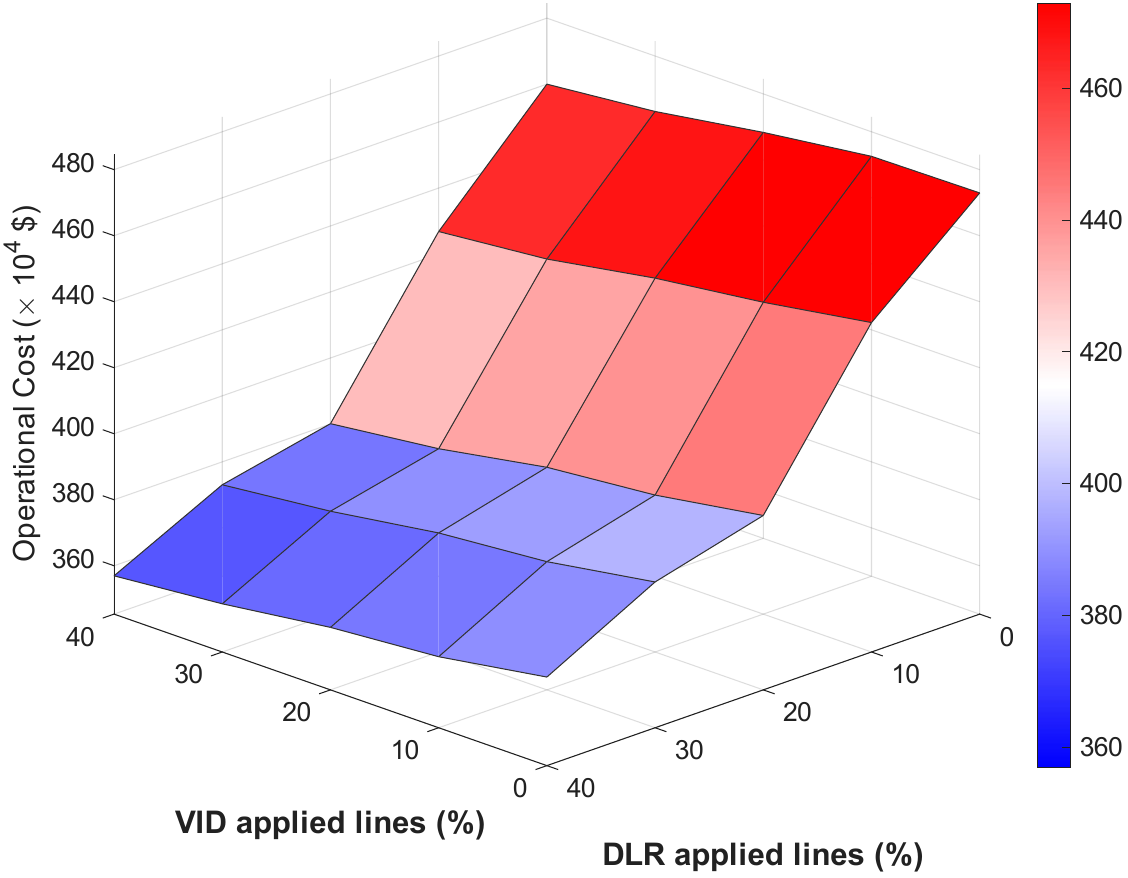}
    \caption{Operational cost surface under different DLR and VID deployment levels (fixed topology, \textsf{Nonlinear}).}
    \label{FIG:3}
\end{figure}

Table~\ref{tab:cost_matrix_nonlinear} reports the operational cost ($C^{\text{Obj}}$), under different DLR and VID deployment levels. Each row corresponds to a DLR deployment level, while each column corresponds to a VID deployment level. The trend is also shown visually in Fig.~\ref{FIG:3}.
Based on Table~\ref{tab:cost_matrix_nonlinear}, the following observations can be made.

\begin{table*}[htbp!]
\centering
\scriptsize
\renewcommand{\arraystretch}{1.6}
\setlength{\tabcolsep}{12.5pt}
\caption{Operational cost matrix under different DLR and VID deployment levels (fixed topology, \textsf{Nonlinear}). Values are reported in $10^4$~\$, and parentheses indicate percentage changes relative to the base case without DLR or VID deployment.}
\label{tab:cost_matrix_nonlinear}
\begin{tabular}{ccccccc}
\toprule
\multicolumn{1}{c}{\textbf{DLR applied lines (\%)}} & \multicolumn{5}{c}{\textbf{VID  applied lines (\%)}} \\
\cmidrule(lr){2-6}
 & 0\%  & 10\%  & 20\%  & 30\%  & 40\%  \\
\midrule
0\%  & 472.95 & 472.56 ($-0.08\%$) & 468.34 ($-0.97\%$) & 463.20 ($-2.06\%$) & 460.00 ($-2.74\%$) \\
\hline
10\% & 445.14 ($-5.88\%$) & 439.83 ($-7.00\%$) & 435.66 ($-7.88\%$) & 429.97 ($-9.09\%$) & 426.82 ($-9.75\%$) \\
\hline
20\% & 398.18 ($-15.81\%$) & 392.89 ($-16.93\%$) & 389.83 ($-17.57\%$) & 383.98 ($-18.81\%$) & 380.09 ($-19.63\%$) \\
\hline
30\%  & 389.51 ($-17.64\%$) & 384.19 ($-18.77\%$) & 381.41 ($-19.36\%$) & 376.52 ($-20.39\%$) & 373.06 ($-21.12\%$) \\
\hline
40\%  & 372.18 ($-21.31\%$) & 366.87 ($-22.43\%$) & 364.28 ($-22.98\%$) & 359.95 ($-23.89\%$) & 356.95 ($-24.53\%$) \\
\bottomrule
\end{tabular}
\end{table*}

\begin{enumerate}[label=(\roman*)]
    \item DLR provides a significantly greater reduction in operational cost compared to VID. At 40\% deployment, DLR alone reduces the cost by 21.31\%, whereas VID alone achieves only a 2.74\% reduction, indicating the higher primary effectiveness of DLR in expanding network capacity.
    
    \item Increasing DLR penetration results in a consistent and substantial decrease in operational cost across all deployment levels. In contrast, the operational cost shows a relatively marginal response to changes in VID penetration, implying that the independent impact of VID on overall cost reduction is limited.
    
    \item The combined deployment of DLR and VID achieves a total cost reduction of up to 24.53\%, confirming a clear synergistic effect. Notably, the cost-saving efficacy of VID is enhanced when paired with DLR; for instance, the absolute cost reduction brought by 40\% VID increases from $12.95 \times 10^4$ \$ under 0\% DLR to $15.23 \times 10^4$ \$ under 40\% DLR. This implies that VID maximizes its routing flexibility when paired with the expanded line capacities provided by DLR.
\end{enumerate}



\begin{figure}
    \centering
    \includegraphics[width=\linewidth]{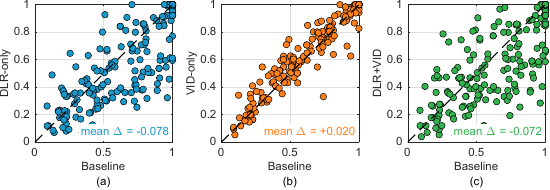}
    \vspace{-1.5em}
    \caption{Comparison of mean line loading under fixed topology: baseline versus (a) DLR-only, (b) VID-only, and (c) DLR+VID. Each dot represents a transmission line, and the axes show line loading in p.u. The 45$^\circ$ line indicates no change.}
    \label{fig:line_loading} 
\end{figure}

We also evaluate how DLR and VID affect the utilization of network resources, which can contribute to lower operational costs through more economical dispatch and reduced load shedding. Figure~\ref{fig:line_loading} compares the mean line loading between the baseline case without DLR/VID and three modified cases: DLR-only, VID-only, and DLR+VID, where each technology is installed on the top priority lines. For each transmission line $l$, we compute the one-day average loading as
$\bar{u}_{l}=\frac{1}{T}\sum_{t=1}^{T}|P_l^{t}|/P_l^{\max, t}$.

With DLR, the dynamic rating $P_l^{\max,t}$ increases according to wind speed and ambient temperature. As a result, most points in the DLR-only case move below the $45^\circ$ reference line, indicating lower mean line loading than in the baseline case. The system-wide average loading decreases from 0.653 to 0.575, corresponding to $\mathrm{mean}\,\Delta\bar{u}=-0.078$. This confirms that DLR relieves line congestion by increasing available transfer capability, which enables more economical generation dispatch and reduces load shedding.

In contrast, VID changes the power flow distribution by modifying line reactance rather than directly increasing line ratings. Therefore, the VID-only case shows only minor deviations from the baseline, with a small average change of $\mathrm{mean}\,\Delta\bar{u}=+0.020$. This indicates that VID primarily redistributes flows across the network, with only a marginal impact on the system-wide average line loading.

When DLR and VID are applied together, the average change is $\mathrm{mean}\,\Delta\bar{u}=-0.072$, which is close to that of the DLR-only case. This suggests that the overall loading relief is mainly driven by DLR, while VID provides additional flow redistribution effects. Overall, the results show that DLR is effective in reducing line loading through dynamic capacity enhancement, whereas VID contributes by reshaping the spatial distribution of power flows.

\begin{figure}
    \centering
    \includegraphics[width=1.0\linewidth]{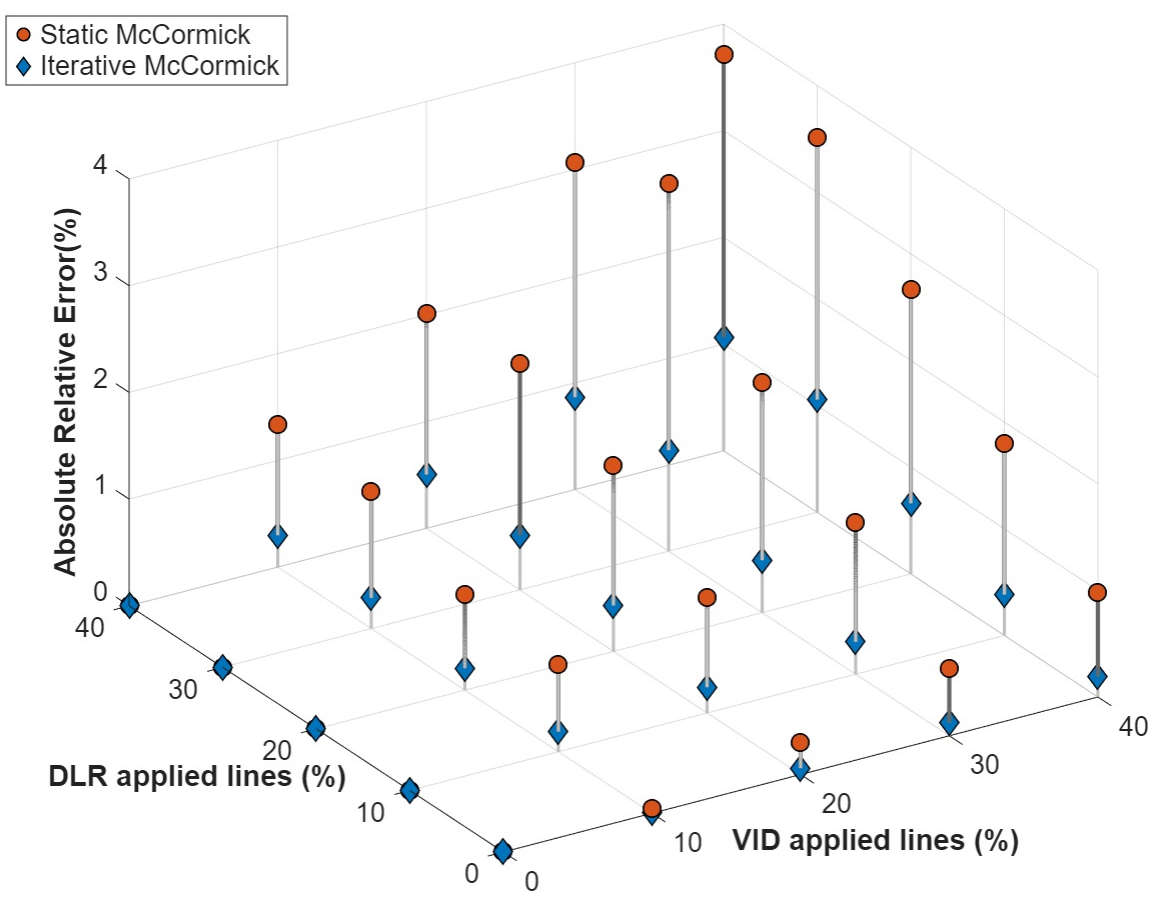}
    \caption{Relative errors of objective values with respect to the \textsf{Nonlinear} solution under fixed topology: \textsf{Iterative McCormick} (blue dots) and \textsf{Static McCormick} (orange dots).}
	\label{FIG:relative_error}
\end{figure}

While the \textsf{Nonlinear} approach provides exact solutions, directly handling the bilinear terms often increases computational burden. To evaluate this, we compare the computational performance of \textsf{Nonlinear}, \textsf{Static McCormick}, and the proposed \textsf{Iterative McCormick} across the 25 DLR--VID deployment scenarios shown in Table~\ref{tab:cost_matrix_nonlinear}. This corresponds to 75 optimization runs in total, and the resulting computation times are summarized in Table~\ref{tab:comp_time_fixed}.
As it can be observed, McCormick-based method achieve more than two orders of of-magnitude reduction in computational time compared to the Nonlinear solver. 
The \textsf{Iterative McCormick} takes more time compared to \textsf{Static McCormick} because of the two iterative corrections (e.g., using 5\% step sizes) to achieve superior solution quality. Nevertheless, is still on the same order of static method.
In terms of accuracy, we compare the solutions using \textsf{Static McCormick} and \textsf{Iterative McCormick} in Figure~\ref{FIG:relative_error}. The relative errors are then calculated with respect to the \textsf{Nonlinear} method.
We can observe that iterative approach achieves lower relative errors compared to the static method, hence achieves closer solution to the nonlinear method.

\begin{table}[htbp]
\centering
\caption{Computational Time Comparison for Fixed Topology (25 Scenarios)}
\label{tab:comp_time_fixed}
\footnotesize
\renewcommand{\arraystretch}{1.3}
\setlength{\tabcolsep}{2pt}

\begin{tabular}{l ccc}
\toprule
\multirow{2}{*}{\textbf{Solution Methodology}} & \multicolumn{3}{c}{\textbf{Computational Time (s)}} \\
\cmidrule{2-4}
 & \textbf{Mean} & \textbf{Std. Dev.} & \textbf{Range [Min, Max]} \\
\midrule
\textsf{Nonlinear}             & \textbf{18.693} & 48.134 & [0.312, 227.434] \\
\textsf{Static McCormick}      & \textbf{0.376}  & 0.190  & [0.312, 0.473]   \\
\textsf{Iterative McCormick}   & \textbf{0.686}  & 0.037  & [0.312, 0.914]   \\
\bottomrule
\end{tabular}
\end{table}






\subsubsection{Optimized Topology through NTO}

In the optimized-topology scenario, the optimization problem in \eqref{eq:optimization_problem_final} is solved by allowing a flexible network configuration, where the binary indicators $h_l^t$, $h_b^t$, and $h_g^t$ are treated as decision variables rather than fixed parameters. Table~\ref{tab:comparison_nto_nonnto_reduction} summarizes the economic benefits of this topology flexibility by comparing the fixed-topology and optimized-topology results. The fixed-topology results are obtained from the \textsf{Nonlinear} reference solution, whereas the optimized-topology results are obtained using \textsf{Iterative McCormick}. 


\begin{table}[!h]
\centering
\caption{Operational Cost and Savings via NTO Under Different Topology Flexibility Settings}
\label{tab:comparison_nto_nonnto_reduction}
\footnotesize
\renewcommand{\arraystretch}{1.1}
\begin{tabular}{cc l cc}
\toprule
\multicolumn{2}{c}{\textbf{Scenario (\%)}} & \multirow{2}{*}{\textbf{Topology}} & \multirow{2}{*}{\textbf{\makecell[c]{Operating\\Cost ($10^4$~\$)}}} & \multirow{2}{*}{\textbf{\makecell[c]{Cost\\Reduction (\%)}}} \\
\cmidrule(lr){1-2}
\textbf{DLR} & \textbf{VID} & \textbf{Setting} & & \\
\midrule
\multirow{3}{*}{0} & \multirow{3}{*}{0}
 & Fixed       & 472.95 & \text{—} \\
 &                   & NTO ($3,1$) & 465.85 & 1.50 \\
 &                   & NTO ($5,2$) & 461.95 & 2.33 \\
\midrule
\multirow{3}{*}{0} & \multirow{3}{*}{20}
 & Fixed       & 468.34 & \text{—} \\
 &                   & NTO ($3,1$) & 461.56 & 1.45 \\
 &                   & NTO ($5,2$) & 458.79 & 2.04 \\
\midrule
\multirow{3}{*}{20} & \multirow{3}{*}{0}
 & Fixed       & 398.18 & \text{—} \\
 &                   & NTO ($3,1$) & 391.09 & 1.78 \\
 &                   & NTO ($5,2$) & 384.15 & 3.52 \\
\midrule
\multirow{3}{*}{20} & \multirow{3}{*}{20}
 & Fixed       & 389.83 & \text{—} \\
 &                   & NTO ($3,1$) & 379.81 & 2.57 \\
 &                   & NTO ($5,2$) & 374.87 & 3.84 \\
\bottomrule
\end{tabular}
\begin{flushleft}
    \scriptsize
    \textbf{Note:} NTO ($l,b$) denotes $\overline{N}_l=l$ and $\overline{N}_{b}=b$.
\end{flushleft}
\end{table}

Based on these comparative results, the following observations can be made: 
   In the base-case scenario, where neither DLR nor VID is deployed, enabling NTO improves operational efficiency. The NTO ($5,2$) setup reduces operating costs by 2.33\% compared to the fixed-topology. 
    In the ($20,20$) scenario, 20\% of the line are installed with DLR and VID. We observed that NTO in this case ($5,2$) achieves cost reduction of 3.84\%. 

Finally, we compare the computation times of \textsf{Nonlinear} and \textsf{Iterative McCormick} to assess the computational tractability of the proposed solution strategy under the optimized-topology setting, as shown in Table~\ref{tab:comp_performance_detail}, which clearly proves superiority of the \textsf{iterative McCormick} approach. As shown in Table~\ref{tab:comp_performance_detail}, the \textsf{Nonlinear} approach failed to reach the target optimality gap within the 3,600s time limit in several scenarios, underlining its limitations for real-time applications, whereas \textsf{iterative McCormick} always achieves the target optimality gap.

\begin{table}[t]
\centering
\caption{Computational performance comparison}
\label{tab:comp_performance_detail}
\footnotesize
\renewcommand{\arraystretch}{1.2}
\setlength{\tabcolsep}{3pt}
\newcolumntype{P}{>{\centering\arraybackslash}p{40pt}}
\newcolumntype{S}{>{\centering\arraybackslash}p{27pt}}
\newcolumntype{M}{>{\centering\arraybackslash}p{58pt}}
\begin{tabular}{@{} P S S M M @{}}
\toprule
\multirow{2}{*}{\makecell[c]{\textbf{Network}\\\textbf{Setting}}}
& \multicolumn{2}{c}{\textbf{Scenario (Line \%)}}
& \textbf{Nonlinear}
& \makecell[c]{\textbf{\textsf{Iterative}}\\\textbf{\textsf{McCormick}}} \\
\cmidrule(lr){2-3} \cmidrule(lr){4-4} \cmidrule(lr){5-5}
& \textbf{DLR}
& \textbf{VID}
& \textbf{(s)}
& \textbf{(s)} \\
\midrule
\multirow{4}{*}{\makecell[c]{$\overline{N}_{l}=3$\\$\overline{N}_{b}=1$}}
& \multirow{2}{*}{0}
  & 0  & \multicolumn{2}{c}{761.9} \\ 
& & 20 & 2491.5 & 1350.8 \\
\cmidrule(lr){2-5}
& \multirow{2}{*}{20}
  & 0  & \multicolumn{2}{c}{645.6} \\ 
& & 20 & 3600\,(1.13\%) & 1339.2 \\
\midrule
\multirow{4}{*}{\makecell[c]{$\overline{N}_{l}=5$\\$\overline{N}_{b}=2$}}
& \multirow{2}{*}{0}
  & 0  & \multicolumn{2}{c}{1688.6} \\ 
& & 20 & 3600\,(1.02\%) & 900.0 \\
\cmidrule(lr){2-5}
& \multirow{2}{*}{20}
  & 0  & \multicolumn{2}{c}{819.2} \\ 
& & 20 & 3600\,(1.68\%) & 1302.8 \\
\bottomrule
\end{tabular}
\begin{flushleft}
    \scriptsize
    \textbf{Note:} Values in parentheses indicate the optimality gap at the 3,600s limit. For scenarios with 0\% VID deployment, the \textsf{Nonlinear} and \textsf{Iterative McCormick} approaches reduce to an identical formulation, and their common computation times are merged.
\end{flushleft}
\end{table}

\subsection{Validation for Other Benchmark Test Cases}
\label{sec:validation_other_cases}

We further validate the scalability and effectiveness of the proposed framework using additional transmission test systems of different sizes: \texttt{case24\_ieee\_rts}, \texttt{pglib\_case588}, and \texttt{case1354pegase}. The corresponding results are summarized in Table~\ref{tab:cost_ls_cases}. The \texttt{case24\_ieee\_rts} system is evaluated over the full 24-hour profile, while \texttt{pglib\_case588} and \texttt{case1354pegase} are evaluated for a representative single time period. For the NTO case, the topology optimization parameters are set to $\overline{N}_{l}=3$ and $\overline{N}_{b}=1$ for all test systems. 

As shown in Table~\ref{tab:cost_ls_cases}, NTO consistently reduces the operational cost compared with the fixed-topology case across all evaluated networks and scenarios. In addition, the combined deployment of DLR and VID generally leads to the lowest operational cost, indicating that these technologies can provide complementary benefits by increasing transmission flexibility and optimizing the network configuration. The relative effectiveness of DLR and VID varies across systems, suggesting that the value of each technology depends on network topology, congestion patterns, and device placement.

The computational performance of the proposed approach is also assessed using the NTO computation times reported in Table~\ref{tab:cost_ls_cases}. For \texttt{case24\_ieee\_rts}, the full 24-hour NTO problem is solved within a few minutes. For the large-scale test systems, where NTO is typically applied to selected operating periods rather than over the entire daily horizon, the proposed \textsf{Iterative McCormick} approach obtains solutions within 30 minutes even for the \texttt{case1354pegase} system. These results demonstrate the scalability and practical applicability of the proposed framework for benchmark transmission systems with different network sizes.

\begin{table}[htbp]
\centering
\caption{Comparison of operational cost and NTO computation time under different test cases}
\label{tab:cost_ls_cases}
\renewcommand{\arraystretch}{1.3}
\resizebox{\columnwidth}{!}{
\begin{tabular}{l c c c c c} 
\toprule
\multirow{2}{*}{\textbf{Test Case}} & \multicolumn{2}{c}{\textbf{Scenario (\%)}} & \multicolumn{2}{c}{\textbf{Operational Cost ($10^4$~\$)}} & \textbf{NTO Time} \\
\cmidrule(lr){2-3} \cmidrule(lr){4-5} \cmidrule(lr){6-6}
& \textbf{DLR} & \textbf{VID} & \textbf{Fixed} & \textbf{NTO} & \textbf{(s)} \\
\midrule
\multirow{4}{*}{\texttt{case24\_ieee\_rts}} 
& 0 & 0   & 843.23 & 820.66 & 131.9 \\
& 0 & 20  & 830.60 & 806.70 & 268.6 \\
& 20 & 0  & 832.34 & 809.86 & 133.8 \\
& 20 & 20 & 811.79 & 788.40 & 248.8 \\
\midrule
\multirow{4}{*}{\texttt{pglib\_case588}$^\dagger$} 
& 0 & 0 & 485.37 & 476.36 & 36.5 \\
& 0 & 20 & 483.07 & 474.26 &  195.6 \\
& 20 & 0 & 454.17 & 446.24 & 30.3 \\
& 20 & 20 & 447.74 & 440.00 & 170.2 \\
\midrule
\multirow{4}{*}{\texttt{case1354pegase}$^\dagger$}
& 0 & 0 & 6,506.4 & 6,474.9 & 935.9 \\
& 0 & 20 & 6,459.7 & 6,430.7 & 1644.4 \\
& 20 & 0 & 6,320.7 & 6,267.9 & 1462.8 \\
& 20 & 20 & 6,205.7 & 6,151.9 & 1639.8 \\
\bottomrule
\end{tabular}
} 
\begin{flushleft}
    \scriptsize
    $^\dagger$ Results obtained for a single time period.     The \texttt{case24\_ieee\_rts} results are obtained over the full 24-hour profile.
\end{flushleft}
\end{table}

\section{Conclusion}
\label{sec:conclusion}
This work presented a co-optimization framework that coordinates different grid enhancing technologies, such as network topology optimization (NTO), variable impedance devices (VIDs), and dynamic line rating (DLR) - equipped transmission lines with an objective to minimize the operational cost. The problem included bilinear constraints due to variable impedance in the topology optimization model, making it a mixed integer non-linear problem. 
This work tackled the nonlinearity by using McCormick relaxation method, which was further enhanced by iterative approach.

The proposed method was validated through extensive simulations on multiple benchmark networks. The results demonstrate that the proposed iterative McCormick method achieves a favorable balance between solution accuracy and computational efficiency. Compared with the nonlinear approach, which provides high accuracy at the expense of significantly longer computation times, and the static McCormick method, which is computationally efficient but less accurate, the proposed method offers an effective trade-off between accuracy and runtime. 

Future work will extend this framework considering other controllable resource such as energy storage systems. 
\bibliographystyle{cas-model2-names}

\bibliography{cas-refs}


\end{document}